\documentclass[pra,aps,twocolumn,eqsecnum,showpacs]{revtex4}

\usepackage{dcolumn}
\usepackage{amsmath}
\usepackage{graphics}

\begin{document}

\title{Diffuse light scattering dynamics under conditions of\\
electromagnetically induced transparency}
\author{V.M. Datsyuk, I.M. Sokolov, D.V. Kupriyanov}
\affiliation{Department of Theoretical Physics, State Polytechnic
University, 195251, St.-Petersburg, Russia}%
\email{Kupr@DK11578.spb.edu}%
\author{M.D. Havey}
\affiliation{Department of Physics, Old Dominion University,
Norfolk, VA 23529}%
\email{mhavey@odu.edu}%

\date{\today }

\begin{abstract}
We show that, under conditions of electromagnetically induced
transparency (EIT), a significant portion of the incident probe
pulse can be transferred into Rayleigh and Raman scattering
channels. The light scattered into the Rayleigh channel emerges
from the sample with an EIT time delay.  We show that a proper
description of the probe light propagation in the sample should
include, in the diffusion dynamics, a spin polariton generated by
the two-photon EIT process.  The results have important
implications for studies of weak light localization, and for
manipulation of single and few photon states in ultracold atomic
gases.
\end{abstract}

\maketitle%


\section{Introduction}

External control of the electrodynamic response of atomic systems
has been revolutionized by the merger of the ideas of coherent
population trapping \cite{Harris,Lukin1,Lukin2,Fleischhauer1} with
the techniques and concepts of ultracold atom physics
\cite{Metcalf}. The potential afforded by the combination was
first demonstrated in the remarkable experiments of Hau, \emph{et
al.} \cite{Hau}, in which a coherent light pulse was compressed in
an ultracold gas of sodium atoms, the excitation having a very
small group velocity $\sim$ 17 m/s. Subsequent extensive
theoretical and experimental research has shown that a combined
atomic-photonic quasiparticle excitation is created in the
ultracold medium, and that the properties of the polariton can be
dynamically manipulated through the external qualities of the
light fields used to prepare and probe the medium
\cite{Lukin1,Lukin2,Fleischhauer1,Fleischhauer2}.  Since then,
studies have shown a broad range of physics associated with
coherent manipulation of propagation of electromagnetic waves
\cite{Braje1,Braje2,Liu,Yan,Budker,Kash,Matsko, Xiao}. For
example, Chaneli\`{e}re and co researchers have recently shown, in
a series of beautiful experiments \cite{Chaneliere}, that it is
possible to generate single photon wave packets, and to map them
into polaritons in an ultracold sample of $^{85}$Rb atoms. The
single photon wave packets could be regenerated after a
controllable delay by judicious application of a control
electromagnetic field. The physical processes associated with
manipulation and storage of individual photon wave packets, and
entanglement of the quantum states of the packets with a
propagating one, are critical elements for quantum information
protocols and quantum memory applications.

Among the essential elements needed for practical applications is
quantitative understanding of coherence loss mechanisms in each
stage of a quantum information protocol.  In many studies of
storage and retrieval of single and multiple quantum wave packets
in ultracold gases, the initially created excitation undergoes
decay as the length of the storage time is increased.  In this
paper, we focus quantitatively on one mechanism which can lead to
such a potential loss of fidelity, and some of the surprising
physics that results from these considerations. In particular, we
consider the diffusely (multiply) scattered light that is
necessarily generated as a result of optical excitation with
temporally finite light pulses. In a general case, the associated
finite spectral band width produces diffusely scattered Rayleigh
and Raman modes in the sample.  These modes show quite different
temporal behavior, in comparison with the coherent forward
scattered probe light. In addition, because the Rayleigh scattered
component maintains coherence with the incident probe beam, a new
type of quasiparticle, a diffuse polariton, is created. Finally,
we point out that the physics of light diffusion in a coherent
medium consisting of a thermal (not ultracold) gas of Rb atoms has
been investigated by Matsko, \emph{et al.} \cite{Matsko}.  We
point out that there have been a large number of studies of the
fascinating physics of coherent population trapping; some
representative papers are cited here
\cite{Braje1,Braje2,Liu,Yan,Budker,Kash,Matsko}. A number of
reviews have also appeared on the subjects of EIT and coherent
population trapping, and contemporary applications
\cite{Fleischhauer1,Harris,Lukin2,Marangos,Arimondo}.

In the remainder of this paper, we first consider in more detail
why, in the context of the previous discussion, the diffuse
scattering channels are important. This is followed by a
description and discussion of the complex susceptibility for the
case of a typical lambda configuration attainable on hyperfine
resonance transitions in an ultracold gas of $^{87}$Rb atoms
\cite{Scully}. The scattering channels are then considered, with
particular attention paid to the Green's function for propagation
in the inhomogeneous and optically anisotropic medium.  We then
present our results describing the temporal behavior of the
forward scattered light and the diffusely scattered fluorescence.
We finally show that the coherence of the multiply scattered
Rayleigh mode can be detected through the appearance of the
coherent backscattering effect for the weak probe beam.

\section{Basic assumptions and calculation approach}
\subsection{Why the scattering channels are important}\label{Why}
\begin{figure}[t]
\includegraphics{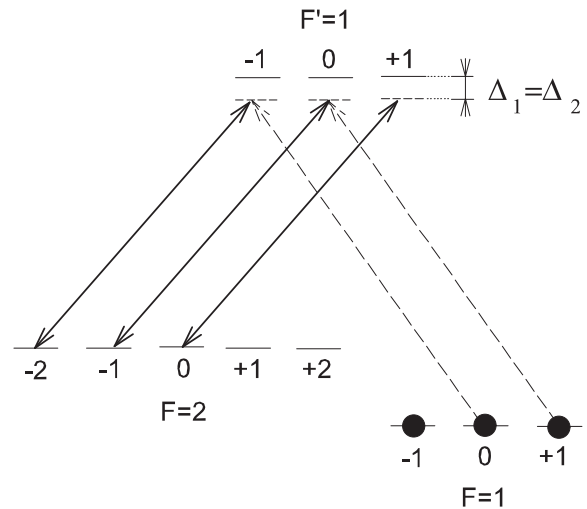}%
\caption{An example of an excitation scheme for observation of the
EIT effect in the system of hyperfine and Zeeman sublevels of the
$D_1$-line of ${}^{87}$Rb. The coupling field is applied with
right-handed circular polarization to the $F=2\to F'=1$ transition
and the probe mode in the orthogonal left-handed polarization
excites the atoms on the $F=1\to F'=1$ transition. The EIT effect
appears for equal detunings of the coupling and probe modes from
atomic resonances: $\Delta_1=\Delta_2$}
\label{fig1}%
\end{figure}

As an example of a system where EIT resonance can be clearly
observed, we consider the $\Lambda$-type configuration in the
hyperfine manifold of the $D_1$-line of ${}^{87}$Rb, see Fig.
\ref{fig1} A strong coupling field with right-handed circular
polarization is applied between the $F=2$ hyperfine sublevel of
the ground state and $F'=1$ hyperfine component of the excited
level. The atoms equally populate the relevant Zeeman states of
the lower hyperfine sublevel $F=1$. In the figure, the detuning of
the coupling (probe) laser from atomic resonance is defined as
$\Delta_1$ ($\Delta_2$). Such an atomic configuration is typical
for EIT observation and can be fashioned in ultracold systems
using for example, magneto-optic or quasistatic dipole traps. Such
configurations have also been widely studied in more traditional
heated atomic gas cells; see for examples, Ref.
\cite{Budker,Kash,Matsko}.

In the present paper, we assume that atoms are so deeply slowed by
an atomic cooling process that we can neglect all effects
associated with atomic motion. These are ideal conditions for EIT
and the probe mode, applied with any polarization to the $F=1\to
F'=1$ hyperfine transition, would find respective $\Lambda$-type
excitation channels, thus generating, as a result of the coherent
Raman process, hyperfine coherence in the ground levels. To set
our terminology, we refer in brief to Rayleigh and elastic Raman
scattering processes as Rayleigh scattering.  The terminology
Raman scattering refers only to hyperfine inelastic Raman
scattering.  All time scales are measured relative to the inverse
lifetime $\Gamma^{-1}$ of the ${}^{87}$Rb resonance transition.

To reliably observe the conversion of the probe pulse into a
polariton-type quasiparticle state, which, as expected, would
forwardly propagate through the sample, some temporal and spectral
requirements should be fulfilled. There are two important temporal
parameters for the process: the pulse duration $\tau_p$ and the
delay time $\tau_d$ between the entrance and emergence of the
probe light pulse. Apparently, for high-fidelity conversion of the
full incident light pulse into the polariton pulse, it is
desirable that $\tau_d>\tau_p$. For the simplest $\Lambda$-type
configuration the delay time can be estimated as
\begin{equation}
\tau_d=\frac{L}{\bar{v}}\sim
n_0\lambdabar^2L\,\frac{\Gamma}{\Omega_R^2}
\label{2.1}%
\end{equation}
where $L$ is the sample length, $\bar{v}$ is a group velocity for
the probe mode at the resonance, $n_0$ is the density of atoms,
$\lambdabar$ is the wavelength divided by $2\pi$ for the probe
radiation, $\Gamma$ is the natural decay rate for the upper state
and $\Omega_R$ is the Rabi frequency for the coupling mode. The
time $\tau_p$ can be estimated via the time-frequency uncertainty
principle, \emph{i.e.} as $\tau_p\sim 1/\Delta_p$, where
$\Delta_p$ is the spectral width of the pulse. The spectral width
$\Delta_p$ is restricted by the condition that, at the relevant
detuning from the EIT resonance, the optical thickness of the
sample $b(\Delta_2)$ at $\Delta_2=\Delta_p$ would be small enough
and the medium would be transparent. Then in order of magnitude
the pulse duration $\tau_p$ is limited by the inequality
\begin{equation}
\tau_p > \sqrt{n_0\lambdabar^2L}\,\frac{\Gamma}{\Omega_R^2}
\label{2.2}%
\end{equation}
For an optically extended medium with $\sqrt{n_0\lambdabar^2L}\gg
1$ the pulse duration can be made shorter than the delay time. But
in reality such an optically dense sample is rather difficult to
prepare in an experiment with ultracold atoms, where the parameter
$n_0\lambdabar^2L$ is typically close to ten or even less in order
of magnitude.

The problem is more subtle if the EIT channel is adjusted for
transport of a portion of "non-classical" light. As a particular
example, one can imagine a pulse of forwardly propagating squeezed
light. Such states of light can be created with an intra-cavity
optical parametric light source and its spectral properties can be
controlled with the quality factor of the cavity. For such a light
source the outgoing radiation can be properly described by the
model proposed by Collett and Gardiner in Ref. \cite{GrdCll} when
the squeezed ($X_1$) and anti-squeezed ($X_2$) quadrature
components are described by the correlation functions with
different relaxation times, which we respectively denote as
$\tau_1$ and $\tau_2$. In principle, there exists an inequality
between these correlation times $\tau_2>\tau_1$, which can be a
strong constraint for high degrees of squeezing. For reliable
transport of the squeezed light with preservation of its unique
statistical properties it would be necessary that
$\tau_p>\tau_2>\tau_1$. The variation of all these temporal
parameters should be limited both by the longer estimation
(\ref{2.1}) and the shorter estimation (\ref{2.2}). Apparently it
would be more difficult to fulfill the EIT criteria for the
squeezed light than for a pure coherent light source.

However, the atomic sample can be probed with a light pulse with
shorter duration than is given by (\ref{2.2}) and the
EIT-mechanism will work even if a portion of the light pulse is
transported via a non-forward scattering channel. For such an
experimental situation the pulse duration $\tau_p$ can be limited
by the time scale $\Gamma/\Omega_R^2$, given by the inverse
spectral width of the transparency window in the local
susceptibility of the medium. Then the input pulse of the probe
light should be transformed into a polariton-type pulse in the
diffuse mode, which now will fill the sample via a coherent
diffusion (non-forward scattering) process.

\subsection{The macroscopic susceptibility and scattering tensors}
The dynamics of the macroscopic polarization, induced by a probe
radiation pulse, is driven by the dynamical susceptibility of the
medium. Referring to the excitation scheme shown in Fig.
\ref{fig1}, under conditions of EIT-resonance, the susceptibility
of the medium for the probe mode entering the sample in a
particular polarization is generated by two coherently interacting
$\Lambda$-type channels. In the laboratory frame, with the
$z$-axis directed along the coupling beam, the susceptibility
tensor has a diagonal form in the basis of circular polarizations
and can be written as the following sum
\begin{eqnarray}
\chi_{q}^{q'}(\mathbf{r},\Delta_2)&=&-\delta_{q}^{q'}\sum_{n(m),m'(m),m}\frac{1}{\hbar}%
\frac{|(\mathbf{d}\mathbf{e}_q^*)_{nm}|^2}{\Delta_2+i\Gamma/2}\,%
\rho_{mm}(\mathbf{r})%
\nonumber\\%
&&\hspace{-0.8 in}%
\times\left\{1-\frac{|V_{nm'}|^2}{\Delta_2+i\Gamma/2}\,%
\frac{1}{\Delta_1-\Delta_2+\Sigma_{nm'}(\Delta_2)}\right\}\,%
\label{3.1}%
\end{eqnarray}
We use standard co/contravariant notation for the basis vectors of
circular polarizations, see \cite{VMK}, which can be expressed by
Cartesian basis vectors as $\mathbf{e}_0=\mathbf{e}_z$,
$\mathbf{e}_{\pm 1}=\mp(\mathbf{e}_x\pm i\mathbf{e}_y)/\sqrt{2}$
\cite{Footnote}.

The first line in (\ref{3.1}) has an isotropic form and describes
the local macroscopic susceptibility in the normal approach of
linear electrodynamics. Here the squared transition dipole moments
$(\mathbf{d}\mathbf{e}_q^*)_{nm}$ between the lower
$|m\rangle\equiv|F,m\rangle$ and upper
$|n\rangle\equiv|F',n\rangle$ Zeeman sublevels are weighted with
the population components of the atomic density matrix
$\rho_{mm}(\mathbf{r})=n_0(\mathbf{r})/(2F+1)$, where
$n_0(\mathbf{r})$ is the local density of atoms at a spatial point
$\mathbf{r}$. In a environment characteristic of ultracold and
trapped atoms, the density distribution is typically inhomogeneous
and the density matrix as well as the susceptibility tensor are
spatially dependent. The frequency detuning $\Delta_2$ is the
offset of the probe mode $\omega_2$ from the resonance
$\Delta_2=\omega_2-\omega_{F'F}$ with $F=1,F'=1$.

The second line in (\ref{3.1}) reveals the contribution of the
EIT-effect. Here $V_{nm'}$ are the transition matrix elements for
the coupling mode between those quantum states
$|n\rangle\equiv|F',n\rangle$ and $|m'\rangle\equiv|F,m'\rangle$
which are subsequently chained with an initial state $|m\rangle$
via the respective $\Lambda$-type excitation channel. This is
indicated in the sum by the dependence of the subscript indices on
$m$: $m'=m'(m)$ and $n=n(m)$. The frequency detuning $\Delta_1$ is
the offset of the coupling mode $\omega_1$ from the resonance
$\Delta_1=\omega_1-\omega_{F'F}$ with $F=2,F'=1$. The pole in the
denominator of Eq.(\ref{3.1}) is shifted due to the Autler-Townes
effect and the self-energy correction is given by
\begin{eqnarray}
\Sigma_{nm'}(\Delta_2)&\equiv&\Delta_{nm'}(\Delta_2)-\frac{i}{2}\Gamma_{nm'}(\Delta_2)%
\nonumber\\%
&=&\frac{|V_{nm'}|^2}{\Delta_2+i\Gamma/2}%
\label{3.2}%
\end{eqnarray}%
The susceptibility tensor (\ref{3.1}) describes \textit{an
anisotropic and optically active medium despite the homogeneous
population of the Zeeman sublevels}. This is a direct consequence
of the EIT-effect. Anisotropy comes from the different
$\Lambda$-type transitions activated by differently polarized
probe modes. This effect has a similar physical nature as various
optical anisotropy effects associated with the presence of the
strong coupling field, see \cite{McGDF,CCKP,SRCHAS}.

The scattering process in a medium is conveniently described by
the scattering tensor formalism. This tensor is responsible for
frequency- and polarization-dependent transformation of an
incident electromagnetic plane wave as a result of its scattering
on an isolated atom. Under conditions of EIT control for the mode
incident on atom at frequency $\omega_2$, the scattering tensor is
given by
\begin{eqnarray}
\hat{\alpha}_{pq}^{(m''m)}(\Delta_2)%
&\equiv& {\alpha}_{pq}^{(m''m)}(\Delta_2)|m''\rangle\langle m|%
\nonumber\\%
\nonumber\\%
&&\hspace{-0.8 in}%
=-\sum_{m'(n),n}\frac{1}{\hbar}%
\frac{(d_p)_{m''n}(d_q)_{nm}}{\Delta_2+i\Gamma/2}\,%
|m''\rangle\langle m|
\nonumber\\%
&&\hspace{-0.8 in}%
\times\left\{1-\frac{|V_{nm'}|^2}{\Delta_2+i\Gamma/2}\,%
\frac{1}{\Delta_1-\Delta_2+\Sigma_{nm'}(\Delta_2)}\right\}\,%
\label{3.7}%
\end{eqnarray}
which determines the amplitude of the outgoing wave for either the
elastic or inelastic scattering channel accompanied by transition
of the atom from the state $|m\rangle\equiv |Fm\rangle$ with $F=1$
to the state $|m''\rangle\equiv|Fm''\rangle$ with F=1 (Rayleigh
channel) or $F=2$ (Inelastic Raman channel). The scattering tensor
should obey the following important identity
\begin{eqnarray}
\int\! d\Omega\sum_{m''{\mathbf \varepsilon}'}%
\frac{\omega_2(\omega_2+\omega_{mm''})^{3}}{c^4}%
\left|{\alpha}_{pq}^{(m''m)}(\Delta_2)%
({\mathbf\varepsilon}^{\prime*})^p\varepsilon^q\right|^2&&%
\nonumber\\%
&&\hspace{-2.8 in}=\frac{4\pi\omega_2}{c}\mathrm{Im} \left[{\alpha}_{pq}^{(mm)}(\Delta_2)%
({\mathbf\varepsilon}^{*})^p\varepsilon^q\right]%
\label{3.7'}
\end{eqnarray}
which is an optical theorem reflecting the unitary property of the
scattering process. Here $\varepsilon^q$ and
$({\mathbf\varepsilon}^{\prime*})^p$ are respectively the
contravariant circular components of the polarization vectors for
the incident photon and of the complex conjugated polarization
vector for the scattered photon. The sum is extended over all
possible scattering channels and all possible output
polarizations. After integration over the full scattering angle
the left hand side reproduces the total scattering cross section.

Because of their similar physical nature, the expressions for the
susceptibility tensor (\ref{3.1}), and for the scattering tensor
(\ref{3.7}) are visualized in similar form. The important
difference is that, in the expression (\ref{3.7}), an arbitrary
reference frame can be assumed and any dipole-type transitions
$|m\rangle\to |n\rangle$ can be initiated. At the same time, each
of the upper states $|n\rangle$ is coupled by the strong field
with a certain selected ground state $|m'\rangle$ that is
indicated in the sum as $m'=m'(n)$. This prohibits the light
scattering for the resonance mode when $\Delta_2=\Delta_1$ but
opens the scattering channel for any non-vanishing detuning.
Moreover, under the EIT effect it becomes quite important which
scattering channel (elastic Rayleigh or inelastic Raman) the probe
photon occupies as it is scattered away from its original forward
propagation. An elastically scattered photon still undergoes
coherent coupling and should emerge from the sample with a
relatively long delay. In contrast, in the Raman channel the
photon freely propagates through the sample with nearly the
(vacuum) speed of light because the medium is ideally transparent
in this case.

\subsection{The Green's function}
Since the probe or scattered beam can propagate in any direction,
its free-path transformation can be properly described by the
Green's function formalism. For the transmitted or elastically
scattered light the Green's function can be expressed in the form
of phase integrals
\begin{eqnarray}
D_{q_1}{}^{q_2}({\mathbf r}_1,{\mathbf r}_2,\omega_2)&=&%
-\frac{\hbar}{|{\mathbf r}_1-{\mathbf r}_2|}\,%
e^{i\phi_0({\mathbf r}_1,{\mathbf r}_2)+ik_2|{\mathbf r}_1-{\mathbf r}_2|}%
\nonumber\\%
&&\hspace{-1.2 in}\times\left[\cos(\phi({\mathbf r}_1,{\mathbf r}_2))\,%
\delta_{q_1}{}^{q_2}%
+\,{\rm i}\sin(\phi({\mathbf r}_1,{\mathbf r}_2))\,%
(\vec{n}\hat{\vec{\sigma}})_{q_1}{}^{q_2}\right]%
\label{3.3}%
\end{eqnarray}%
Here $k_2=\omega_2/c$ and $\vec{n}=\vec{n}(\Delta_2)$ is the
"unit" symbolic vector, the components of which are defined below.

The parameters of the phase integrals can be expressed by the
major components of the susceptibility tensor, which are defined
by Eq.(\ref{3.1}) and related to the probe beam either
co-propagating in the forward direction or perpendicular to the
coupling beam. In this sense, expression (\ref{3.3}) gives a
radiation zone asymptote for the retarded-type Green's function
and the polarization indices $q_1,q_2$ are defined in the
reference frame naturally linked with the ray direction, such that
$z\parallel {\mathbf r}_1-{\mathbf r}_2$. In this frame the
Green's function has a non-diagonal form in a natural basis of
local circular polarizations with $q_{1,2}=\pm 1$
(left-hand/right-hand).

The phase integrals are given by
\begin{eqnarray}
\phi_0({\mathbf r}_1,{\mathbf r}_2)&=&\frac{2\pi\omega_2}{c}\,%
\int_{{\mathbf r}_2}^{{\mathbf r}_1}\chi_0({\mathbf r},\Delta_2)\,ds%
\nonumber\\%
\phi({\mathbf r}_1,{\mathbf r}_2)&=&\frac{2\pi\omega_2}{c}\,%
\int_{{\mathbf r}_2}^{{\mathbf r}_1}\chi({\mathbf r},\Delta_2)\,ds%
\label{3.4}%
\end{eqnarray}%
and evaluated along a ray from point $\mathbf{r}_2$ to point
$\mathbf{r}_1$. The integrand $\chi_0({\mathbf r},\Delta_2)$ is
expressed in terms of major-axes components of the susceptibility
tensor as
\begin{eqnarray}
\chi_0({\mathbf r},\Delta_2)&=&\frac{1+\cos^{2}\beta}{4}%
\left[\chi_{-1}^{-1}({\mathbf r},\Delta_2)+%
 \chi_{+1}^{+1}({\mathbf r},\Delta_2)\right]%
\nonumber\\%
&&+\frac{\sin^{2}\beta}{2}\,\chi_{0}^{0}({\mathbf r},\Delta_2)%
\label{3.5}%
\end{eqnarray}%
where $\beta$ is the polar angle between the ray and the coupling
beam directions. The function $\chi({\mathbf r},\Delta_2)$ is
given by
\begin{equation}
\chi({\mathbf r},\Delta_2)=\left[\sum\limits_{\mathrm{i}=1}^{3}%
\chi_{\mathrm{i}}^2({\mathbf r},\Delta_2)\right]^{1/2}%
\label{3.5'}
\end{equation}
where
\begin{eqnarray}
\chi_1({\mathbf r},\Delta_2)&=&\frac{\sin^2\beta}{2}\cos\gamma%
\left\{\frac{1}{2}\left[\chi_{-1}^{-1}({\mathbf r},\Delta_2)+%
 \chi_{+1}^{+1}({\mathbf r},\Delta_2)\right]\right.%
\nonumber\\%
 &&\phantom{\frac{\sin^2\beta}{2}\cos\gamma}%
 \left.\phantom{\frac{1}{2}}-\chi_{0}^{0}({\mathbf r},\Delta_2)\right\}%
\nonumber\\%
\chi_2({\mathbf r},\Delta_2)&=&\frac{\sin^2\beta}{2}\sin\gamma%
\left\{\frac{1}{2}\left[\chi_{-1}^{-1}({\mathbf r},\Delta_2)+%
 \chi_{+1}^{+1}({\mathbf r},\Delta_2)\right]\right.%
\nonumber\\%
 &&\phantom{\frac{\sin^2\beta}{2}\cos\gamma}%
 \left.\phantom{\frac{1}{2}}-\chi_{0}^{0}({\mathbf r},\Delta_2)\right\}%
\nonumber\\%
\chi_3({\mathbf r},\Delta_2)&=&\frac{\cos\beta}{2}%
\left[\chi_{+1}^{+1}({\mathbf r},\Delta_2)-%
 \chi_{-1}^{-1}({\mathbf r},\Delta_2)\right]%
\label{3.6}%
\end{eqnarray}%
Here $\gamma$ is an azimuthal angle indicating the uncertainty in
directions of a pair of orthogonal axes fixing the reference frame
in the plane transverse to the ray. The "unit" symbolic vector
$\vec{n}$ is defined by its components as
\begin{equation}
n_{\mathrm{i}}=n_{\mathrm{i}}(\Delta_2)=\frac{\chi_{\mathrm{i}}({\mathbf r},\Delta_2)}%
{\chi({\mathbf r},\Delta_2)}%
\end{equation}
and is independent of $\mathbf{r}$. It is described in the general
case by the set of complex components $(n_1,n_2,n_3)$.

\section{Results}
\subsection{The forward transmitted pulse}

\begin{figure}[tp]
\includegraphics{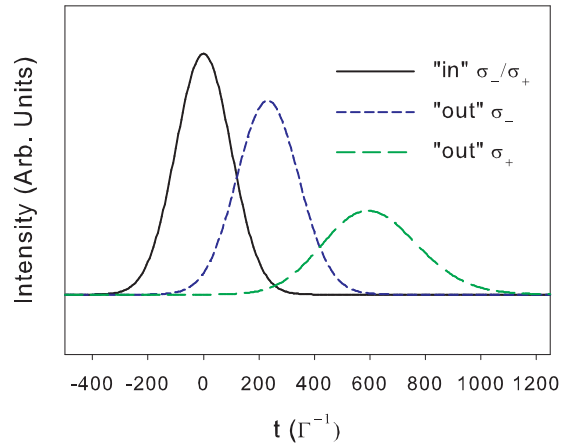}%
\caption{Intensity profiles for light pulses transmitted through a
sample of ${}^{87}$Rb atoms via the EIT channels. The coupling
field is applied on the empty $F=2\leftrightarrow F'=1$ transition
with right-handed circular polarization and the probe field,
exciting the atoms on the $F=1\to F'=1$ transition, is either
left-handed polarized (blue, short-dash curve) or right-handed
polarized (green, longer-dash curve). The black curve indicates
the original Gaussian-type pulse with a duration of
$\tau_p=100\,\Gamma^{-1}$.}
\label{fig2}%
\end{figure}

\begin{figure}[tp]
\includegraphics{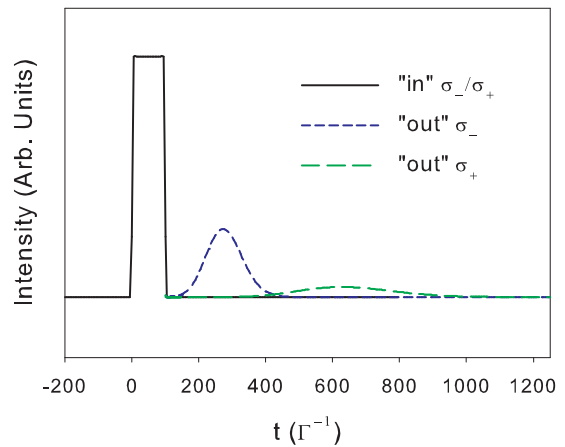}%
\caption{Same as in Fig. \ref{fig2} but for a rectangular input
pulse profile.}
\label{fig3}%
\end{figure}

The graphs in Figs. \ref{fig2} and \ref{fig3} show how the
original pulse profile is modified after propagation through the
sample in an EIT channel. The calculations were done for two
different pulse shapes, these being Gaussian-type (Fig.
\ref{fig2}) and rectangular-type (Fig. \ref{fig3}), with pulse
widths of $\tau_p=100\,\Gamma^{-1}$. In our numerical simulations
we considered as an example ${}^{87}$Rb atoms, when the coupling
field and the probe light excitations were respectively applied at
$F=2\leftrightarrow F'=1$ (empty) and $F=1\to F'=1$ (equally
populated) hyperfine transitions, as shown in Fig. \ref{fig1}. The
coupling field is always right-hand circularly polarized and the
probe field can have varying polarization, being either left- or
right-hand circularly polarized. The calculations were based on
attainable parameters of an ultracold atomic cloud confined in a
magneto optic trap. For a Gaussian atom distribution in the trap,
the weak-field optical depth, on resonance and through the center
of the trap is given by $b_0=\sqrt{2\pi}n_0\sigma_0r_0$, where
$n_0$ is the peak density and $\sigma_0$ is the normal (not
modified by the EIT effect) resonance cross section, $r_0$ is the
radius of the cloud. The dependencies shown in Figs. \ref{fig2}
and \ref{fig3} reproduce the pulse transformation after its
passing through such an atomic sample with the size $r_0=0.25\,cm$
and peak optical depth $b_0=50$. All the parameters are in
accordance with the condition that the delay time $\tau_d$ is
longer than the duration $\tau_p$ of the original pulse. The
inequality $\tau_d>\tau_p$ can be controlled by appropriate choice
of the Rabi frequency for the coupling field
$\Omega_c=2|V_{nm'}|$, which we defined with respect to the
$m'=\{F=2,M=-1\}\leftrightarrow n=\{F'=1,M'=0\}$ hyperfine Zeeman
transition. For the dependencies shown in Figs. \ref{fig2} and
\ref{fig3}, the Rabi frequency is given by $\Omega_c=0.4\Gamma$.

For the case of a Gaussian profile the output pulse preserves the
original shape but becomes more extended in time. The highest
fidelity channel for reproduction of the original pulse is
obtained when the coupling and probe fields are in orthogonal
polarization states (blue curve in Fig.\ref{fig2}). The loss of
the pulse energy is only 10\% for this case. But for the pulse
transmitted through the sample in the right-handed polarization
channel there is up to 40\% energy loss. This is a clear
manifestation of the scattering process, which cannot be ignored,
as we argued by simple estimations in Section \ref{Why}.

The modification of the pulse profile manifests itself even more
dramatically if the probe pulse has a rectangular shape. Due to
dispersion the original pulse shape is completely transformed in
the extended medium to a Gaussian form and the losses of energy
are more significant for this case. That effect is shown by the
calculations of the outgoing pulse profile presented in Fig.
\ref{fig3} for two different polarization channels. The
calculations were done for the initial parameters similar to those
used in the calculation of the dependencies for Fig. \ref{fig2}.
For the outgoing pulses shown in Fig. \ref{fig3} the losses are up
to 60\% for the left-handed and up to 80\% for the right-handed
polarization channels.

\subsection{The scattered pulse}

\begin{figure}[tp]
\includegraphics{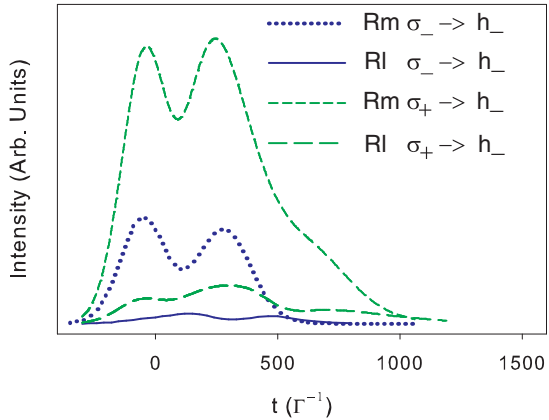}%
\caption{The intensity profiles for the portion of the light pulse
scattered at $90^0$ to the direction of the incident pulse. The
curves represent the Rayleigh (Rl) channel ($F=1\to F'=1\to F=1$)
and the Raman (Rm) channel ($F=1\to F'=1\to F=2$), with the input
polarization state, and polarization channel of the emerging light
as indicated in the caption. In all cases, the observation channel
corresponds to detection of light with left-hand helicity
($h_{-}$).}
\label{fig4}%
\end{figure}

The plots in Fig. \ref{fig4} reproduce the time dependence of
instantaneous intensity for the fractions of the light pulse,
originally incoming with Gaussian profile shown in Fig.
\ref{fig2}, and scattered by the sample in a direction orthogonal
to the incident probe beam. Note that we designate the
polarization of the scattered light in terms of its helicity,
which is a natural choice when considering the radiation
propagating towards a detector.  In Fig. 4, the curves indicate
our results for the Rayleigh scattering channel ($F=1\to F'=1\to
F=1$) and the Raman channel ($F=1\to F'=1\to F=2$). Let us recall
here that for the sake of brevity we do not distinguish between
elastic Raman and Rayleigh channels. The calculational parameters
were chosen the same as for the dependencies of Figs. \ref{fig2}
and \ref{fig3}. The solid and dotted blue curves illustrate the
portion of the scattered light pulse incoming with left-handed
circular polarization ($\sigma_{-}$) and emerging the sample in
the left-hand helicity polarization ($h_{-}$). In turn, the
short-dashed and long-dashed green curves illustrate the portion
of the scattered light pulse incoming with right-handed
polarization ($\sigma_{+}$) and emerging the sample again in the
left-hand helicity polarization ($h_{-}$). Thus, for the intensity
profiles shown in Fig. \ref{fig4}, the outgoing light is always
considered in the left-hand helicity polarization channel.

As follows from these dependencies, the scattered light emerges
the sample with a rather long delay in units of $\Gamma^{-1}$.
This is an indication of the slow light phenomenon typically
associated with the EIT effect, but for the scattered light. The
portion of the input pulse, distorted in the scattering process,
propagates through the sample faster and is more extended in time
than the transmitted pulse. This can be explained by the spectral
structure of the scattered pulse, which has a deficit of near
resonance photons for which the EIT mechanism works ideally. For
the scattered light pulse leaving the sample via a Rayleigh
channel, the group velocity is greater than for the transmitted
incident pulse. The photons created in the Raman channel leave the
sample with the speed of light \emph{in vacuo} because the medium
is fully transparent for them. The spectral notch in the
distribution of the scattered photons leads to the beating effect
in the time behavior of the output light intensity, which is
clearly seen for all the scattering channels presented in Fig.
\ref{fig4}.

To show the difference between the usual incoherent scattering and
scattering in the environment supporting the EIT-effect, let us
turn to the physical background of the process. If the medium is
illuminated by a single mode probe field, which is precisely in
the resonance with the coupling field such that
$\omega_1=\omega_2+\omega_{21}$, where $\omega_{21}$ is the
hyperfine splitting in the diagram of Fig. \ref{fig1}, the entire
atom-field system is described by the following density matrix
\begin{equation}
\hat{\rho}=\frac{1}{3}\sum_{\mathrm{j}=1,2}|\Psi_{\mathrm{j}}\rangle\langle\Psi_{\mathrm{j}}|%
+\ldots%
\label{4.1}
\end{equation}
where dots denote the contribution of non-coupled state and the
sum $\mathrm{j}=1,2$ is extended over two $\Lambda$ schemes shown
in figure \ref{fig1}. The wave functions contributing to the
expansion (\ref{4.1}) are given by
\begin{eqnarray}
|\Psi_{\mathrm{j}}\rangle&=&\left[\frac{\Omega_p^{(\mathrm{j})}}%
{\sqrt{\Omega_p^{(\mathrm{j})2}+\Omega_c^{(\mathrm{j})2}}}|1\rangle_{\mathrm{j}}%
-\frac{\Omega_c^{(\mathrm{j})}}%
{\sqrt{\Omega_p^{(\mathrm{j})2}+\Omega_c^{(\mathrm{j})2}}}|2\rangle_{\mathrm{j}}\right]%
\nonumber\\%
\nonumber\\%
&&\phantom{\frac{\Omega_p^{(\mathrm{j})}}
{\sqrt{\Omega_p^{(\mathrm{j})2}+\Omega_c^{(\mathrm{j})2}}}}%
\times|\mathrm{Field}\rangle%
\label{4.2}%
\end{eqnarray}
where $|1\rangle_{\mathrm{j}}$ and $|2\rangle_{\mathrm{j}}$ are
respectively the left and right atomic spin states coupled with
the j-th $\Lambda$-type excitation channel and
$\Omega_c^{(\mathrm{j})}$ and $\Omega_p^{(\mathrm{j})}$ are the
Rabi frequencies respectively for the coupling and for the probe
fields. For the sake of simplicity we presume both the Rabi
frequencies to be real parameters. It is a crucial point of the
EIT effect that despite the fact that even though the atomic
substate given in the square-brackets of (\ref{4.2}) is not an
eigenstate of the atomic Hamiltonian, the full wavefunction
$|\Psi_{\mathrm{j}}\rangle$ is an eigenstate of the entire
atom-field system if the field is in coherent state. Thus under
conditions of perfect $\Lambda$-resonance the system is not able
to scatter the light.

However if the probe mode is not in exact two-photon resonance the
situation becomes more complicated. In the single atom case and as
a first approximation there will be coherent beats with exchange
of a photon between coupling and probe modes of the field, which
are initiated by low frequency phase oscillation of the atomic
spin coherence. In a macroscopic system these beats are
collectivized and transformed to a polariton wave packet created
in the sample by a pulse of the probe field enveloped by a set of
near resonance spectral modes. For a short distance the polariton
wave propagates in the forward direction but it will be scattered
at long distances if the medium is optically extended. However the
probe field, even being scattered via Rayleigh channels, remains
in a coherent state and can be also enveloped by the modes inside
the EIT spectral window. Thus the polariton wave \textit{does not
disappear} as a result of Rayleigh scattering but
\textit{transforms to a diffuse coherent mode}. The quantum nature
of such a diffuse polariton can be verified if the control field
is switched off and switched on again with delay. Then the
propagation of the polariton wave in the forward direction and via
Rayleigh scattering channel will be stopped and be regenerated
again with a time delay. Thus the scattered light, as well as the
transmitted light, can be stored in the medium and recovered on
demand. That is a unique property of the light propagation under
EIT conditions.

\subsection{The coherent backscattering process}

\begin{figure}[tp]
\includegraphics{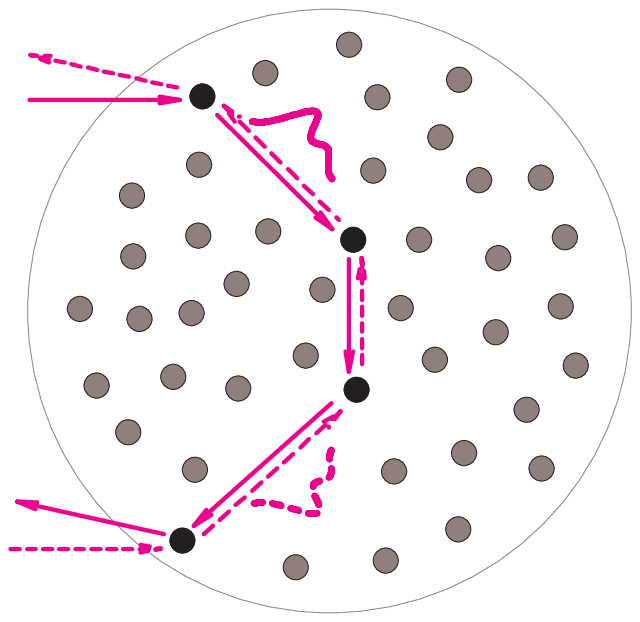}%
\caption{A schematic diagram explaining the coherent
backscattering phenomenon. The polaritonic waves scattered in the
medium follow pairs of reciprocal scattering paths indicated by
the solid and dashed traces.  The emergent electromagnetic field
in the backwards direction can show interference between the
propagating modes following the two paths.}
\label{fig5}%
\end{figure}

A clear indicator of the coherence in the Rayleigh scattering
channel is the appearance of the coherent backscattering (CBS)
process. This effect is a typical example of a "which path"
interference fashioned by a disordered medium. For a portion of a
light wave emerging the sample in the backward direction there is
constructive or, for some special cases, destructive interference
\cite{KSH} between the scattering amplitudes along any multiple
scattering chain. The various manifestations of the CBS phenomenon
in ultracold atomic systems \cite{LTBMMK,Strontium1,KSKSH}, and
under different physical conditions, has been the subject of many
discussions during the last decade, see Kupriyanov, \emph{et al.}
for a review \cite{KSSH}. Some future prospects in this field have
been summarized by Havey and Kupriyanov \cite{HKComment}. In the
context of light scattering in an atomic environment supporting
the EIT effect, the CBS process can be interpreted as interference
between two unknown reciprocal paths for the diffuse polaritonic
wave propagating inside the sample via Rayleigh scattering as
shown schematically in Fig. \ref{fig5}. The specific nature of the
EIT-resonance essentially modifies the time behavior of the
interference component of the outgoing wave packet in comparison
with how it is usually observed for a pulse-type excitation of an
opaque atomic medium.

For the scattering process developing in the standard conditions
of elastic scattering and for a rather long excitation probe pulse
the "ladder" (non-interference) and the "crossed" (interference)
components have an approximately similar dependence on time. The
important difference in their time behavior is mainly observable
in the transient stages of the excitation process, see Ref.
\cite{KSSH}. Inside the pulse the relation between the
instantaneous magnitudes of these components, usually expressed in
terms of the enhancement factor for the CBS output intensity, has
typically no time dependence. The enhancement factor approaches a
stable and constant value inside the pulse. This value reproduces
the enhancement factor for the steady state regime in conditions
of a single mode excitation of the atomic sample with a resonant
probe light. A crucial contrast for the CBS process observable in
an environment of the EIT effect is that the resonant carrier mode
of the light pulse cannot be scattered and the spectral profile
for the pulse fraction, deflected by the medium from its original
forward propagation, always has a narrow gap near this mode. That
results in a quite unusual time behavior for the interference
overlap of two reciprocal fragments of the backscattered wave
packet.

\begin{figure}[tp]
\includegraphics{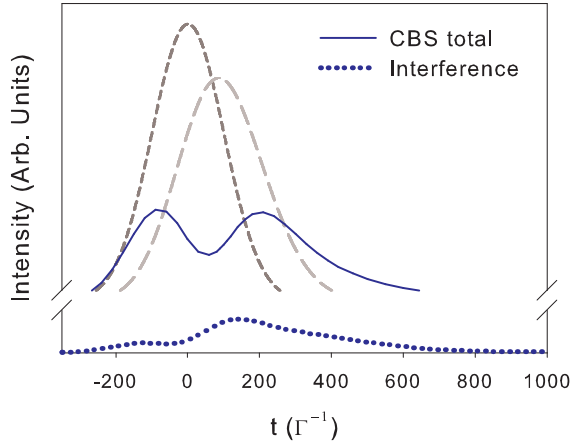}%
\caption{The temporal profiles for the fraction of the light pulse
scattered in the backward direction for the $\sigma_{-}\to h_{-}$
Rayleigh scattering channel. The blue solid curve is the total
intensity profile and the dotted blue curve is the interference
contribution. The short- and long-dashed gray curves respectively
indicate the input and the transmitted pulses. For these graphs
the ordinate only qualitatively reproduces the relative magnitudes
of the pulse intensities.}
\label{fig6}%
\end{figure}

\begin{figure}[tp]
\includegraphics{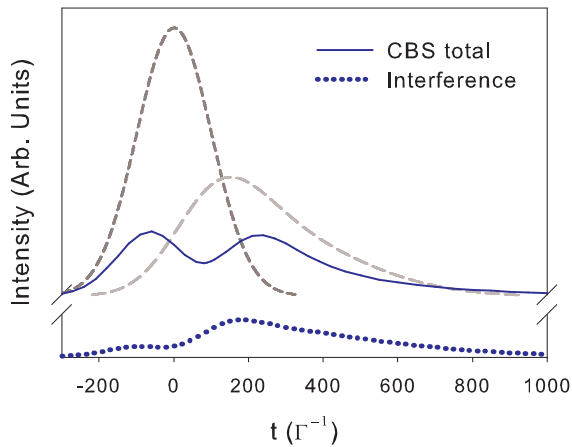}%
\caption{Same as in Fig. \ref{fig6} but for the $\sigma_{+}\to
h_{-}$ Rayleigh scattering channel.}
\label{fig7}%
\end{figure}

The dependencies of Figs. \ref{fig6} and \ref{fig7} illustrate the
temporal behavior of the instantaneous intensity for the fraction
of the probe pulse scattered in the backward direction via the
Rayleigh channels $\sigma_{-}\to h_{-}$ and $\sigma_{+}\to h_{-}$.
The outgoing intensity is mainly contributed by the "ladder"-type
terms such that the interference contribution, shown in the
magnified bottom parts of the graphs, gives actually only a few
percent of the total intensity. For convenience in these figures
the gray curves indicate the time profiles for the incoming and
transmitted pulses. Let us point out that these profiles slightly
differ from those which are shown in Figs. \ref{fig2} and
\ref{fig4} because the aperture of the probe beams are selected
differently for the two calculations. In Fig. \ref{fig2} the cross
section of the beam was less than the cross sectional area of the
atomic cloud, which is a natural requirement for measurements in
slow or stopped light experiments. Alternatively, for observing
the CBS effect from all atoms of the ensemble it is more natural
to make the cross-section of the probe beam bigger than the cross
sectional area of the atomic cloud. The important feature of the
dependencies displayed in Figs. \ref{fig6} and \ref{fig7} is that
the "crossed"-type terms lead to an oscillating interference
enhancement near the midpoint of the backscattered pulse profile.

\begin{figure}[tp]
\includegraphics{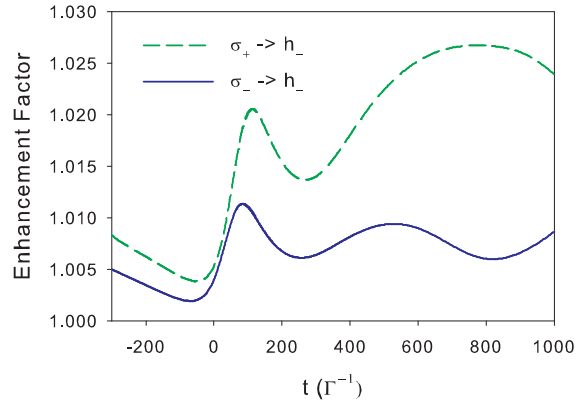}%
\caption{The enhancement factor for $\sigma_{+}\to h_{-}$ (green
dashed curve) and $\sigma_{-}\to h_{-}$ (blue solid curve)
scattering channels.}
\label{fig8}%
\end{figure}

The time behavior of the interference effect can be properly
described in terms of the enhancement factor
\begin{equation}
\eta(t)=\frac{I_S(t)+I_L(t)+I_C(t)}{I_S(t)+I_L(t)}%
\label{4.3}
\end{equation}
where $I_S(t)+I_L(t)$ is the contribution of single scattering and
of the ladder terms in multiple scattering and $I_C(t)$ is the
contribution of the crossed terms. The dependence of $\eta(t)$ on
time is illustrated in the graphs of Fig. \ref{fig8}. It is an
intriguing consequence of the calculation results that the
behavior of the enhancement factor is generally quite complicated
because of the nontrivial manifestation of different spectral
components of the input pulse in formation of the backscattered
fraction of the outgoing pulse in different scattering orders. The
most important seems an oscillating enhancement near the midpoint
of the scattered pulse, which corresponds with the polaritonic
wave packet overlap inside the medium.

\section{Conclusions}
In summary, we have made a theoretical and calculational
investigation of the Rayleigh and Raman scattered light modes in a
medium configured to demonstrate electromagnetically induced
transparency in the coherently forward scattered light.  In
particular, for an anisotropic and inhomogeneous sample of
ultracold atoms we have examined the time evolution of the forward
scattered light, the diffusely (multiply) scattered light in a
characteristic right-angle fluorescence geometry, and in the
backscattering configuration used to study weak localization of
light in ultracold atomic gases.  The intensity of the forward
scattered light is modified by both Rayleigh and Raman scattering
of Fourier components of the incident pulse which lie outside the
EIT transmission window. In particular, the intensity of a light
pulse regenerated in a typical 'stopped-light' experiment is
significantly modified by the loss of these components, which in
part maintain their coherence with respect to the incident light.
In addition, these components produce complex time dependence in
both the fluorescence and in the coherent backscattering
geometries. The interferences responsible for the coherent
backscattering enhancement are a clear indicator of the coherence
of the scattering processes.  We emphasize the fundamental
conclusion that light scattered into the Rayleigh modes maintains
its coherence with respect to the original incident pulse.  This
is manifested by interferometric enhancement of the backwards
scattered intensity from the sample. This in turn implies that,
along with the polariton associated with the forward scattered
light, a diffuse and coherently related polariton coexists within
the volume of the ultracold atomic sample. We close by noting that
a deeper understanding of the mesoscopic properties of the
diffusive excitation is clearly required, and this report
represents a first step in this direction.  Further theoretical
and experimental research into this area is underway.

\section*{Acknowledgments}
We appreciate the financial support of the National Science
Foundation (NSF-PHY-0355024), the Russian Foundation for Basic
Research (RFBR-05-02-16172-a), INTAS (Project ID: 7904) and the
North Atlantic Treaty Organization (PST-CLG-978468). D.V.K. would
like to acknowledge financial support from the Delzell Foundation,
Inc.


\begin{thebibliography}{99}
\bibitem{Footnote} It is important to recognize that in
this complex basis set the contravariant basis-vectors are given
by $\mathbf{e}^q=\mathbf{e}_q^*$ and the metric tensor has
non-diagonal form: $(\mathbf{e}_q)_p=(-)^q\delta_{q\,-p}$ such
that a right hand polarization vector with $q=+1$ has negative
covariant projection with $p=-1$.
\end{thebibliography}
\end{document}